\documentclass{PoS}

\usepackage{amsmath}

\title{Chiral Analysis of the Generalized Form Factors of the Nucleon}

\ShortTitle{Chiral Analysis of the GFFs of the Nucleon}

\author{\speaker{Marina Dorati}$^{a}$, Tobias A. Gail $^{b}$ and Thomas R. Hemmert$^{b}$
         \\
\llap{$^a$}Dipartimento di Fisica Nucleare e Teorica, Universita' degli Studi di Pavia and INFN\\     I-27100 Pavia, Italy\\
\llap{$^b$}Physik-Department, Theoretische Physik T39, TU M\"unchen \\D-85747 Garching, Germany\\

E-mail: \\
\email{marina.dorati@pv.infn.it}\\
        \email{tgail@ph.tum.de}\\
        \email{themmert@ph.tum.de}}

\abstract{We apply the methods of Chiral Perturbation Theory to the analysis of the first moments of the Generalized Parton Distributions in a Nucleon, usually known as \textit{generalized form factors}. These quantities are currently also under investigation in Lattice QCD analyses of
baryon structure, providing simulation results at large quark masses to be extrapolated to the "real world" via Chiral Effective Field Theory.
We have performed a leading-one-loop calculation in the covariant framework of Baryon Chiral Perturbation Theory (BChPT), predicting both the momentum and the quark-mass dependence for all the vector and axial (generalized) form factors. In particular we discuss the results for the
limit of vanishing four-momentum transfer where the GPD-moments reduce to
the well known moments of Parton Distribution Functions (PDFs). We fit our
 results to available lattice QCD data, extrapolating down to the physical point. We conclude by presenting outstanding results from a combined fit to different GPDs-moments. }

\FullConference{The XXV International Symposium on lattice Field Theory\\
                 July 30 - August 4 2007\\
                 Regensburg, Germany}

\begin{document}

\def\sl#1{\slash{\hspace{-0.2 truecm}#1}}
\def\g{g_{A}}
\def\F{F_{\pi}}
\def\m{m_\pi}
\def\M{M_0}
\def\I{I_{11}}
\def\Mn{M_N}
\def\tM{\tilde{M}}
\def\t{\tilde}

\section{Introduction}
In this talk we discuss the findings of references \cite{DGH} and
\cite{DHaxial} where the generalized form factors of the nucleon
were analyzed in the framework of covariant Baryon Chiral
Perturbation Theory (BChPT). In standard SU(2) BChPT the results for
form factors typically depend on two variables, the momentum
transfer squared and the quark-mass and on a number of low energy
constants (LECs). In this work we study the quark-mass dependence of
the isoscalar- and isovector-vector as well as the isovector-axial
generalized form factor $A_{2,0}(t)$ at zero momentum transfer and
predict the physical value of those quantities by determining
previously unknown LECs by a fit of the BChPT results to lattice QCD
data.
Working in twist-2 approximation, the parity-even part of the structure of the nucleon
is encoded via two Generalized Parton Distribution functions (GPDs)
$\textit{H}^{q}(x,\xi,t)$ and $\textit{E}^{q}(x,\xi,t)$ \cite{reviews}. 

Moments of GPDs can be interpreted
much easier and are connected to well-established hadron structure observables. E.g. the zero-th order (Mellin-) moments in the variable $x$
correspond
to the contribution of quark $q$ to the well known Dirac and Pauli form factors $F_1(t),\,F_2(t)$ of the nucleon:
\begin{equation}
\int_{-1}^{1}dx\,x^0\,\textit{H}^{q}(x,\xi,t)=F_{1}^{q}(t)\:\:,\:\:
\int_{-1}^{1}dx\,x^0\,\textit{E}^{q}(x,\xi,t)=F_{2}^{q}(t).
\end{equation}
Our aim is the application of the methods of ChPT to the analysis of the {\em first} moments in $x$ of these nucleon GPDs
\begin{equation}
\int_{-1}^{1}dx\,x\,\textit{H}^{q}(x,\xi,t)=A_{2,0}^{q}(t)+(-2\xi)^{2}\textit{C}_{2,0}^{q}(t)\:\:, \:\:
\int_{-1}^{1}dx\,x\,\textit{E}^{q}(x,\xi,t)=B_{2,0}^{q}(t)-(-2\xi)^{2}\textit{C}_{2,0}^{q}(t)
\end{equation}

where one encounters {\em three generalized form factors}
$A_{2,0}^q(t),\,B_{2,0}^q(t),\,C_{2,0}^q(t)$ of the nucleon for each
quark flavor $q$. For the case of 2 light flavours the {\em
generalized isoscalar (u+d) and isovector (u-d) form factors} have
been already studied in a series of papers at leading-one-loop order
in the non-relativistic framework of Heavy Baryon ChPT (HBChPT)
\cite{GPDs}. We will provide the first analysis of these generalized
form factors utilizing the methods of covariant BChPT for 2 light
flavors \cite{GSS}. Our BChPT formalism \cite{GH} makes use of a
variant of Infrared Regularization \cite{BL} for the loop diagrams
and is constructed in such a way that we {\em exactly} reproduce the
corresponding HBChPT result of the same chiral order in the limit of
small pion masses. For the complete analytical expressions of the
discussed results and a detailed description of the formalism used
for the calculation we refer to \cite{DGH} and \cite{DHaxial}.

\section{The Generalized Form Factors of the Nucleon in ChPT}

In ChPT one can direcly access the isoscalar (\textit{s}) and
isovector (\textit{v}) contribution to the generalized form factors
of the nucleon by evaluating the following matrix elements
\cite{reviews}:
\begin{eqnarray}
 i\langle p'\vert\overline{q}\gamma_{\{ \mu}\overleftrightarrow{\textit{D}}_{\nu\}}\,q\vert p\rangle_{u+d}
&=& \overline{u}(p') \bigg[{\textit{A}_{2,0}^s (\Delta^2)}\gamma_{\{\mu}\overline{p}_{\nu\}}
- \frac{{\textit{B}_{2,0}^s (\Delta^2)} }{2M_{N}}\:\Delta^{\alpha}i\sigma_{\alpha\{\mu}\overline{p}_{\nu\}}
+\frac{ {\textit{C}_{2,0}^s (\Delta^2)} }{M_{N}}\:\Delta_{\{\mu}\Delta_{\nu\}}\bigg]\, \frac{{\bf 1}}{2}\, u(p), \nonumber \\
& & \label{defs}\\
i\langle p'\vert\overline{q}\gamma_{\{ \mu}\overleftrightarrow{\textit{D}}_{\nu\}}\,q\vert p\rangle_{u-d}
&=& \overline{u}(p') \bigg[{\textit{A}_{2,0}^v (\Delta^2)}\gamma_{\{\mu}\overline{p}_{\nu\}}
- \frac{{\textit{B}_{2,0}^v (\Delta^2)} }{2M_{N}}\:\Delta^{\alpha}i\sigma_{\alpha\{\mu}\overline{p}_{\nu\}}
+\frac{ {\textit{C}_{2,0}^v (\Delta^2)} }{M_{N}}\:\Delta_{\{\mu}\Delta_{\nu\}}\bigg]\, \frac{\tau^a}{2}\, u(p). \nonumber \\
& & \label{defv}
\end{eqnarray}

\begin{figure}
\begin{center}
\hspace{-2cm}
\includegraphics[width=.55\textwidth]{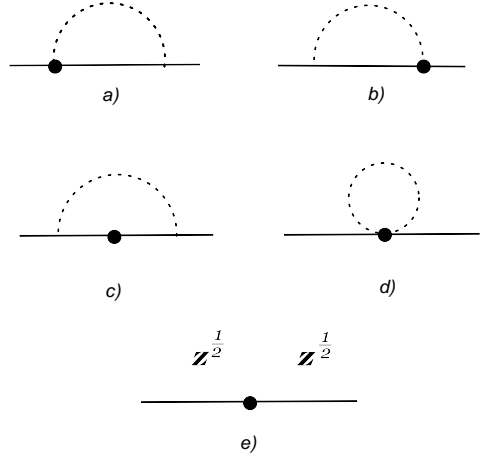}
\caption{Loop diagrams contributing to the first moments of the GPDs of a nucleon at leading-one-loop order in BChPT. The solid and dashed lines represent nucleon and pion propagators respectively. The solid dot denotes a coupling to an external tensor field.}
\label{loop}
\end{center}
\end{figure}

The brackets $\{ \ldots \}$ denote the completely symmetrized and traceless combination of all indices in an operator. $u\,(\overline{u})$ is
a Dirac spinor of the incoming (outgoing) nucleon of mass $M_N$, for which the quark matrix-element is evaluated. 

From a powercounting analysis we find that the Feynman diagrams
contributing to the first moments of GPDs of a nucleon at
leading-one-loop order [$\mathcal{O}(p^2)$] in covariant ChPT are
the ones depicted in Fig.\ref{loop}.

\section{Analysis of the results for $A_{2,0}^{(v,s)}(t=0)$}

In this section we present an analysis of the generalized isovector-
and isoscalar-vector form factors $A_{2,0}^{v,s}(t)$ in the forward
limit $t\rightarrow 0$. The details of the ChPT calculation as well
as a complete analysis of the form factors $B_{2,0}^{v,s}(t)$,
$C_{2,0}^{v,s}(t)$ and their connection to the spin physics sector
can be found in ref\cite{DGH}.

In the forward limit $t\rightarrow 0$ the generalized form factors $A_{2,0}^{s,v}(t=0)$ can be understood as moments
of the ordinary Parton Distribution Functions (PDFs) $q(x), \,\bar{q}(x)$ \cite{reviews}:
\begin{eqnarray}
\langle x\rangle_{u\pm d}&=&A_{2,0}^{s,v}(t=0)=\int_0^1 dx\,x\left(q(x)+\bar{q}(x)\right)_{u\pm d}.
\end{eqnarray}
Experimental results are available for $\langle x\rangle$ in proton- and ``neutron-'' targets, from which one can estimate the isoscalar and isovector
quark contributions at the physical point \cite{xproton} at a regularization scale $\mu$. We choose $\mu=2$ GeV for our
comparisons with phenomenology.

For the PDF-moment $A_{2,0}^v(t=0)$
we obtain to ${\cal O}(p^2)$ in BChPT
\begin{eqnarray}
A_{2,0}^{v}(0) &\equiv& \langle x\rangle_{u-d} \nonumber \\
    &=&a_{2,0}^v+\frac{a_{2,0}^v\m^2}{(4\pi F_{\pi})^2}\Bigg\{-(3\g^2+1)\log\frac{\m^2}{\lambda^2}-2\g^2
    +\g^2\frac{\m^2}{\M^2}\bigg(1+3\log\frac{\m^2}{\M^2}\bigg) \nonumber \\
    &&  -\frac{1}{2}\g^2
    \frac{\m^4}{\M^4}\log\frac{\m^2}{\M^2}
    +\g^2 \frac{\m}{\sqrt{4\M^2-\m^2}}\bigg(14-8\frac{\m^2}{\M^2}+\frac{\m^4}{\M^4}\bigg)\arccos
    {\left(\frac{\m}{2\M}\right)}\Bigg\} \nonumber \\
    & &+\frac{\Delta a_{2,0}^v\g\m^2}{3(4\pi \F)^2}\Bigg\{2\frac{\m^2}{\M^2}\bigg(1+3\log\frac{\m^2}{\M^2}\bigg)-\frac{\m^4}{\M^4}\log\frac{\m^2}{\M^2}
    +\frac{2\m(4\M^2-\m^2)^{\frac{3}{2}}}{\M^4}\arccos\bigg(\frac{\m}{2\M}\bigg)\Bigg\} \nonumber \\
    & &+4\m^2 \frac{c_8^{(r)}(\lambda)}{M_0^2} + {\cal O}(p^3). \label{A20v0}
\end{eqnarray}
\begin{figure}
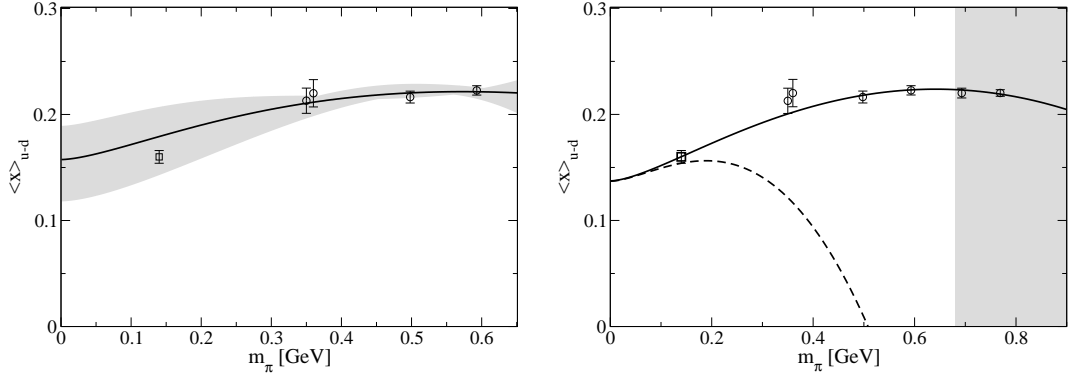

\begin{center}
\includegraphics[width=.45\textwidth]{x1.eps}\hspace{.5cm}\includegraphics[width=.45\textwidth]{x2.eps}
\caption{\textbf{Left panel}. Fit I of the $\mathcal{O}(p^2)$ result
of Eq.(3.2)
 to the LHPC lattice data of ref.\cite{LHPC07}. \textbf{Right panel}.
 Fit II of the $\mathcal{O}(p^2)$ BChPT result of Eq.(3.2)
  to the LHPC lattice data of ref.\cite{LHPC07} {\em and} to the physical point (solid line).
  The dashed curve shown corresponds to $\mathcal{O}(p^2)$ result in the HBChPT truncation.}
\label{x}
\end{center}
\end{figure}
\begin{table}[tb]
\begin{center}
\begin{tabular}{|c|c|c|c|}
\hline
 & $a_{2,0}^v$ & $\Delta a_{2,0}^v$ & $c_8^r$(1GeV)
\\
\hline
Fit I (4 points - 2 parameter) & 0.157 $\pm$ 0.006 &  0.210 (fixed)& -0.283 $\pm$ 0.011  \\
Fit II (6+1 points - 3 parameter) & 0.141 $\pm$ 0.0057 & 0.144 $\pm$ 0.034 & -0.213 $\pm$ 0.03 \\
\hline
\end{tabular}\\
\caption{Values of the couplings resulting from the two fits to the LHPC lattice data for $\langle x \rangle_{u-d}$ \cite{DGH}. The errors shown are only
statistical and do neither include uncertainties from possible higher order corrections in ChEFT nor from systematic uncertainties connected with the
lattice simulation.} \label{table}
\end{center}
\end{table}


Most LECs in this expression are well known from analyses
of chiral extrapolation functions
\cite{PMWHW}. However, the sizes of  $a_{2,0}^v$,
$\Delta a_{2,0}^v$ and  $c_8^{(r)}(\lambda)$ are only poorly known at this point.
The coupling $\Delta a_{2,0}^v$ is related to the spin-dependent analogue of the mean momentum fraction, namely  $\langle \Delta x\rangle_{u-d}$ (see Section 4 and \cite{DHaxial}).
In a first fit (Fit I) to lattice data we constrain $\Delta a_{2,0}^v$ from the
phenomenological value of $\langle \Delta x\rangle_{u-d}^{phen.}\approx 0.21$
and perform a 2-parameter fit with the couplings
$a_{2,0}^v,\,c_8^{(r)}(1\mbox{GeV})$ at the regularization scale $\lambda=1$ GeV. We fit to the LHPC  lattice data for this quantity as
given in ref.\cite{LHPC07}, including lattice data up to effective pion
masses of $m_{\pi} \approx$ 600 MeV. The resulting values for the fit parameters together with their statistical
errors are given in table \ref{table} and the resulting chiral extrapolation function is shown as the solid line in the left hand side of Fig.\ref{x}.
The
extrapolation curve tends towards smaller values for small quark-masses, but does not quite reach the phenomenological value at the
physical point, which is {\em not} included in the fit. Since
BChPT is based on a systematic perturbative expansion, it does not
only provide us with the result at a certain order, but also allows for
an estimate of possible higher order effects.
From dimensional analysis we
know that the leading chiral contribution to $\langle x\rangle_{u-d}$ beyond our
calculation takes the form   ${\cal O}(p^3)\sim\delta_A
\,\frac{m_\pi^3}{\Lambda_\chi^2 M_0}+...$. Constraining
$\delta_A$ between values $-1,\ldots,+1$ (the natural scale of all couplings in the observables considered here is below 1) and
repeating the fit with this uncertainty term included leads to the grey band
indicated in Fig.\ref{x}.
As one can see the phenomenological value for $\langle x\rangle_{u-d}$ lies well within that band of possible next-order corrections, giving us
{\em no indication} that something may be inconsistent with the large values for $\langle x\rangle_{u-d}$ typically found in lattice QCD
 simulations for large quark-masses.

As stated in the Introduction, the covariant BChPT scheme used in
this analysis is able to reproduce exactly the corresponding
non-relativistic HBChPT result at the same order by the appropriate
truncation in $1/(16\pi^2F_\pi^2M_0)$. In order to also compare the
${\cal O}(p^2)$ HBChPT result of refs.\cite{AS} with the ${\cal
O}(p^2)$ covariant BChPT result of Eq.(\ref{A20v0}) we perform a
second fit (Fit II): We fit the covariant expression for $\langle
x\rangle_{u-d}$ of Eq.(\ref{A20v0}) again to the LHPC lattice data
and we constrain the coupling $\Delta a_{2,0}^v$ in such a way, that
the resulting chiral extrapolation curve reproduces the
phenomenological value of $\langle x\rangle_{u-d}^{phen.}=0.160 \pm
0.006$ \cite{xproton} exactly for physical quark masses. The
parameter values for this Fit II are again given in table
\ref{table}, whereas the resulting chiral extrapolation is shown as
the solid line in the right hand side of Fig.\ref{x}. We would like
to emphasize that the curve looks very reasonable, connecting the
physical point with the lattice data of the LHPC collaboration in a
smooth fashion. For the comparison with HBChPT we now utilize the
very same values for $a_{2,0}^v$ and $c_8^{(r)}$ of Fit II. The
resulting curve based on the ${\cal O}(p^2)$ HBChPT truncation is
shown as the dashed curve in Fig.\ref{x}. One observes that this
leading-one-loop HBChPT expression agrees with the covariant result
between the chiral limit and the physical point, but is not able to
extrapolate on towards the lattice data.

We therefore conclude that the smooth extrapolation behaviour of the covariant ${\cal O}(p^2)$ BChPT expression for $\langle x\rangle_{u-d}$
of Eq.(\ref{A20v0}) between the chiral limit and the region of present lattice QCD data is due to an {\em infinite tower} of
$\left(\frac{m_\pi}{M_0}\right)^n$ terms.

\begin{figure}
\begin{center}
\includegraphics[width=.45\textwidth]{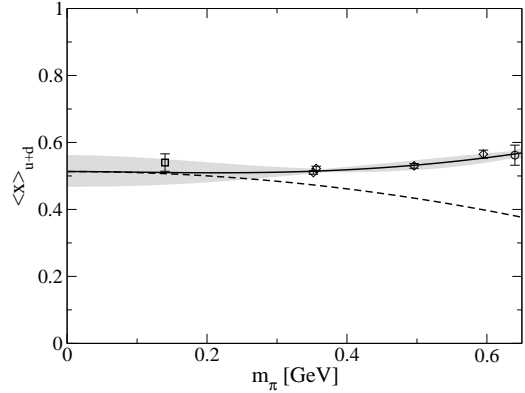}
\caption{Two-parameters Fit of the $\mathcal{O}(p^2)$ BChPT result of ref.{\cite{DGH}} to the
 LHPC lattice data of ref.\cite{LHPC07} and to the QCDSF data of ref.\cite{QCDSF_dat}.
  We obtain  $a_{20}^s=0.513\pm0.006$ and $ c_9=-0.064\pm0.005$ as values for the free parameters.
  The band shown indicate estimate of higher order possible corrections.
   The dashed line correspond to the respective HBChPT results at this order.}
\label{xs}
\end{center}
\end{figure}

The same analysis can also be performed for the isoscalar generalized
form factor $A_{2,0}^{s}(t)$ which in the forward limit reduces to $A_{2,0}^{s}(0)=\langle x\rangle_{u+d}$.
Fig.\ref{xs} shows a 2-parameter fit of the $\mathcal{O}(p^2)$
covariant BChPT results for this observable \cite{DGH} to
LHPC and QCDSF data of reference \cite{LHPC07} and \cite{QCDSF_dat}.
We want to stress that the experimental value of the quantity $\langle
x\rangle_{u+d}$ is not included in the fit.
As the plot shows, the obtained chiral extrapolation curve is
very satisfying, consistently linking
 the lattice data at large quark-masses with the phenomenological
value.
In contrast, the correspondent result in the heavy baryon limit represented by the
dashed line in Fig.\ref{xs} not even allows for an interpolation
between the presently available lattice data and results from experiments.


\section{Combined Fit}

We have extended the analysis to the first moments of the axial GPDs\hspace{.1cm} $\t{H}^{q}(x,\xi,t)$ and $\t{E}^{q}(x,\xi,t)$
\begin{equation}
\int_{-1}^{1}dx\,x\,\tilde{H}^{q}(x,\xi,t)=\tilde{A}_{2,0}^{q}(t)\:\:,\:\:
\int_{-1}^{1}dx\,x\,\tilde{E}^{q}(x,\xi,t)=\tilde{B}_{2,0}^{q}(t).
\end{equation}
Again, in the limit of vanishing
four-momentum transfer the isovector form factor
$\tilde{A}_{2,0}^{v}(t\rightarrow 0)$ is directly connected to the
spin dependent analogue of the mean momentum fraction $\langle\Delta x\rangle_{u-d}$
\begin{eqnarray}
\tilde{A}_{2,0}^v (0)\,&=\,\langle \Delta x\rangle_{u-d}=
\int_{0}^{1}
dx\:\:x\,(q_{\downarrow}(x)-q_{\uparrow}(x))\big\vert_{u-d}\\\nonumber
 &=\,\Delta a_{2,0}^v+
\mathcal{O}(p^2)
\end{eqnarray}
where $\Delta a_{2,0}^v$ corresponds to the chiral limit value of
$\langle \Delta x\rangle_{u-d}$.

\begin{figure}
\begin{center}
\includegraphics[width=.5\textwidth]{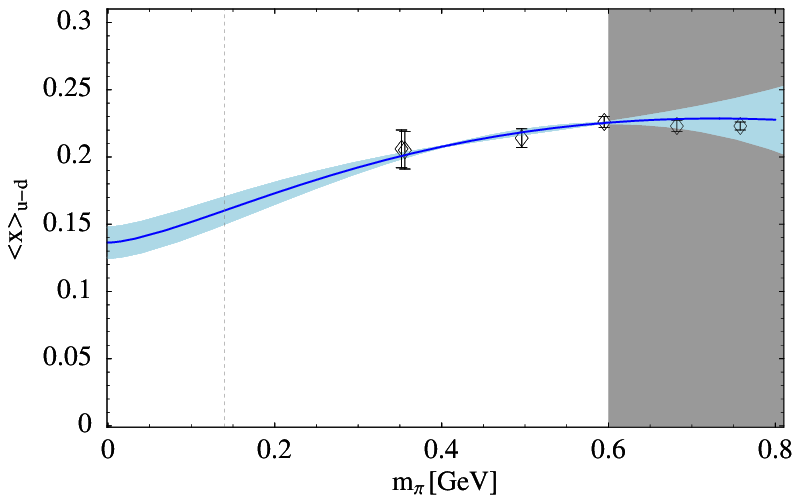}\includegraphics[width=.5\textwidth]{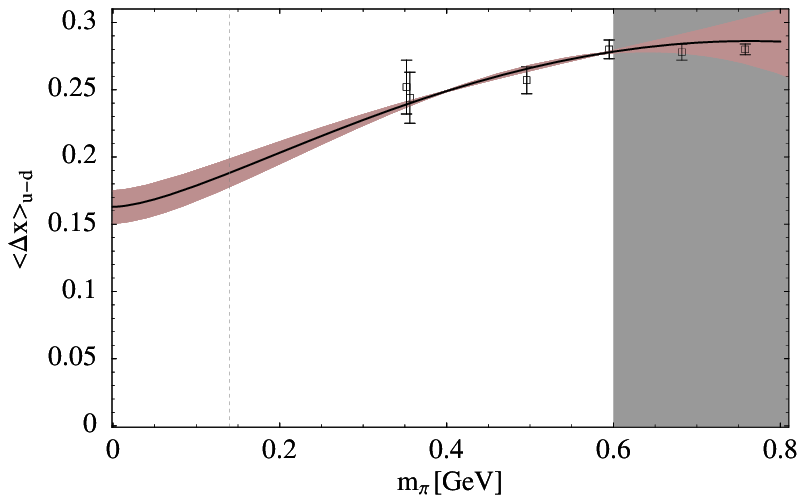}
\caption{Combined FIT of the $\mathcal{O}(p^2)$ results of ref.
\cite{DHaxial} to the lattice data of ref. \cite{LHPC07}. Note that
the phenomenological values at physical pion mass were not included
in the fit. The bands shown indicate estimate of higher order possible corrections.}
\label{combfit}
\end{center}
\end{figure}

Looking at the $\mathcal{O}(p^2)$ BChPT expression for $\langle \Delta x\rangle_{u-d}$ \cite{DHaxial} one can easily observe that each isovector moment ($\langle x\rangle_{u-d}$ and $\langle \Delta x\rangle_{u-d}$) depends on 3 unknown
parameters: 2 couplings ($a_{2,0}^v$, $\Delta a_{2,0}^v$) and one
counterterm. As the same couplings contribute in  both moments, it is
hoped that a simultaneous fit of our BChPT results to the lattice data of ref.\cite{LHPC07} can considerably reduce the statistical errors.
 As one can
see from Fig.\ref{combfit}, the results of this procedure are pretty
outstanding, given that the values at the physical pion mass were
not included in the fit! The chiral curvature in both observables
naturally bends down to the phenomenological value for lighter quark
masses, leading to a very satisfactory extrapolation curve.

We would like to stress that this efficient cross-talk between the ChPT results for $\langle
x\rangle_{u-d}$ and $\langle \Delta
x\rangle_{u-d}$  occurs only in the
covariant framework, while in the non-relativistic approach both
observables are completely independent at this order.

We conclude that combined fits of several observables characterized
by a common subset of ChEFT couplings are the winning strategy
towards the most reliable chiral extrapolations of lattice QCD results.

\end{document}